\documentclass[12pt]{article}

\usepackage{a4wide}
\usepackage[pdftex,usenames,dvipsnames]{color}
\usepackage{graphics}
\usepackage{amsfonts}

\newcommand{\nit}{\noindent}

\newcommand{\np}{\newpage}
\newcommand{\dsp}{\displaystyle}
\newcommand{\vs}[1]{\vspace{#1 ex}}
\newcommand{\hs}[1]{\hspace{#1 em}}
\newcommand{\bfr}{\begin{flushright}}
\newcommand{\efr}{\end{flushright}}
\newcommand{\bc}{\begin{center}}
\newcommand{\ec}{\end{center}}
\newcommand{\ben}{\begin{enumerate}}
\newcommand{\een}{\end{enumerate}}

\newcommand{\be}{\begin{equation}}
\newcommand{\ee}{\end{equation}}
\newcommand{\ba}{\begin{array}}
\newcommand{\ea}{\end{array}}
\newcommand{\ct}{\cite}
\newcommand{\bit}{\bibitem}
\newcommand{\dd}[2]{\frac{\partial{#1}}{\partial{#2}}}
\newcommand{\ag}{\alpha}

\newcommand{\del}{\delta}
\newcommand{\eps}{\epsilon}
\newcommand{\ve}{\varepsilon}

\newcommand{\sg}{\sigma}

\newcommand{\vf}{\varphi}
\newcommand{\og}{\omega}
\newcommand{\Gam}{\Gamma}

\newcommand{\Fg}{\Phi}

\newcommand{\Og}{\Omega}

\newcommand{\bfq}{\bold{q}}

\newcommand{\bfP}{\bold {P}}

\newcommand{\bFg}{\mbox{\boldmath{$\Fg$}}}

\newcommand{\un}{\underline{n}} 
\newcommand{\up}{\underline{p}}

\newcommand{\ueps}{\underline{\eps}}

\newcommand{\uvf}{\underline{\vf}}

\newcommand{\cD}{{\cal D}}
\newcommand{\cH}{{\cal H}}

\newcommand{\lh}{\left(}
\newcommand{\rh}{\right)}
\newcommand{\ld}{\left.}

\begin{document}

\pagestyle{empty}

\bfr
NIKHEF/2013-027
\efr
\vs{7}

\bc
{\large {\bf Time-reparametrization invariance}} \\
\vs{3}

{\large {\bf  and Hamilton-Jacobi approach to the cosmological $\sg$-model}}
\vs{10}

{\large J.W.\ van Holten$^1$ and R.\ Kerner$^2$}
\vs{5}

 $^1 \; \; $ NIKHEF, Science Park 105, Amsterdam
\vskip 0.1cm
and Lorentz Institute, Leiden University, Leiden, Netherlands.
\vskip 0.1cm

E-mail: v.holten@nikhef.nl
\vskip 0.2cm
$^2 \; \;$ 
LPTMC, Universit\'e Paris-VI - CNRS UMR 7600 , \\ Tour 24, 4-\`eme , Boite 121,
4 Place Jussieu, 75005 Paris, France 
\vskip 0.1cm
E-mail: richard.kerner@upmc.fr 
\vskip 0.6cm

August 13, 2013
\vskip 0.7cm
{\small
{\bf Abstract} }
\ec
\vs{1}
\nit
{\small
The construction of physical models with local time-reparametrization invariance is reviewed. Negative-energy 
contributions to the hamiltonian are shown to be crucial for the realization of this reparametrization symmetry.  
The covariant formulation of the dynamics is used to develop a time and gauge invariant Hamilton-Jacobi theory.  
This formalism is applied to solve for the cosmology of a homogeneous universe of the  
Friedmann-Lemaitre-Robertson-Walker type. After a discussion of empty universes, the FLRW theory is 
extended with homogeneous scalar fields generically described by a $\sg$-model on some scalar manifold. 
An explicit gauge-invariant solution is constructed for the non-linear $O(N)$-models. 
}
\vfill


\np 
~ \hfill

\np
\pagestyle{plain}
\pagenumbering{arabic}

\section{Introduction} 

Reparametrization invariance is an essential ingredient of many fundamental theories of physics, 
including particle models, general relativity, supergravity and string theory \ct{einstein1923, 
deser-zumino1976a, brink-divecchia-howe1976,ferrara-etal1976,deser-zumino1976b,polyakov1981}. 
An important consequence of this large local symmetry is the imposition of constraints, especially the 
hamiltonian constraint restricting the dynamics to a fixed hypersurface in phase space, characterized 
by the vanishing of the hamiltonian \ct{adm1962,wheeler1962,dewitt1967a}. 

A local realization of such constraints usually requires the introduction of unphysical degrees of 
freedom, such as gauge fields, ghosts and Lagrange multipliers, which later have to be removed 
from the theory to exhibit the relevant relations between physical quantities predicted by the model
\ct{feynman1962,dewitt1967b,faddeev-popov1967}. Especially in quantum field theory there is an 
intricate set of rules for dealing with these difficulties, including the techniques of gauge fixing, ghost
fields and BRST-cohomology \ct{henneaux-teitelboim1992,vholten2002}.   

In this paper we consider a relatively simple, but physically relevant class of theories with 
time-reparametrization invariance, where the dynamics can be followed in complete detail, while 
circumpassing almost entirely the difficulties mentioned above. These models describe the evolution 
of the homogeneous background space-time and scalar fields in models of cosmology in the context 
of the Friedmann-Lemaitre-Robertson-Walker metric (FLRW). For an overview see ref.\ \ct{weinberg2008};
a recent exposition of the dynamics of the homogeneous backgrounds is found in \ct{vholten2013a}. 
In this paper the main tool for our treatment of the dynamics of these extended mini-superspaces is 
the Hamilton-Jacobi method \ct{mtw1970,peres1962,padmanabhan1989,padmanabhan-sing1990}. 
In effect, for theories in this class we can construct Hamilton-Jacobi functions in terms of physical 
degrees only, without having to specify a particular choice of gauge. 

We consider in particular FLRW cosmologies extended with light scalar fields whose dynamics 
is governed by the geometry of the target manifold, i.e.\ $\sg$-models like the non-linear 
$O(N)$-model \ct{frieman-etal1995,cadamuro-redondo2011}. It turns out, that in these models 
the evolution of the cosmological scale factor is almost independent of the specifics of the scalar 
theory, except for the energy density it creates. At the same time, the dynamics of the scalar fields 
can be solved in detail for the various cosmological scenarios. Apart from the potential relevance for 
cosmology ---applications to theories of inflation \ct{guth2007,linde2004} or quintessence 
\ct{wetterich1995,zlatec-wang-steinhardt1999,armendariz-picon-etal2001,hebecker-wetterich2001}--- 
this provides one with a laboratory for developing a detailed understanding of the interplay 
between scalar field dynamics and cosmological evolution of a spatially homogeneous universe. 
Actually, our solutions might also be used in principle as background field solutions for theories 
of density fluctuations and gravitational waves, like those used e.g.\ in the analysis of the 
CMB radiation \ct{mukhanov-etal1991,linde-mukhanov-sasaki2005,martin-ringeval-vennin2013}.

\section{Time reparametrization invariance} 

In this section we review the implementation and consequences of time-reparametrization 
invariance for dynamical systems with a finite number of (reduced) degrees of freedom, 
such as in particle mechanics or the cosmology of homogeneous universes. Such systems
are described by a set of configuration variables $\bfq = (q^i, ..., q^r)$; for simplicity we 
assume these variables to take values in the real numbers. Being concerned at this point
with classical systems, we also take it for granted that these values can not jump; more 
precisely, we take the evolution of the system $\bfq$ to be represented by a continuous and 
differentiable curve in the $r$-dimensional configuration space. This curve is a map from the 
real line into the configuration space, such that each point on the curve is labeled by a real 
number $t$, interpreted as a time variable. Any such curve $\bfq(t)$ describes a possible 
history of the system. 

There exist systems the dynamics of which is independent of the choice of time variable. 
Such systems often possess an internal degree of freedom acting as a clock, and any
choice of time parameter merely serves to parametrize the points of a history, allowing to
relate them to the value of the internal clock variable \ct{misner1972,kuchar1981,unruh-wald1989}.  
Examples will be given below. 

First, we set up the formalism for dealing with time reparametrization invariance in 
the frame work of lagrangean dynamics. We consider standard lagrangeans, functions
of the configuration variables $\bfq(t)$ and the velocities $\dot{\bfq}(t)$. If the configuration
is to be independent of the choice of time variable, a redefinition of time $t \rightarrow t'(t)$
must leave the values of $\bfq$ at the same point in history unchanged:
\be
\bfq'(t') = \bfq(t).
\label{1.1}
\ee
Thus the variables $q^i(t)$ are scalars (0-forms) on the time line. Time-invariant integration
is achieved by introducing a 1-form $N(t) dt$, where $N$ is the lapse function, transforming
under reparametrizations as
\be
N'(t') dt' = N(t) dt.
\label{1.2}
\ee
The inverse lapse function can then be used to define covariant time derivatives:
\be
\cD \bfq = \frac{1}{N}\, \dot{\bfq}, \hs{2} \lh \cD \bfq \rh'(t') = \cD \bfq(t).
\label{1.3}
\ee
This provides a construction for invariant lagrangeans $L(\bfq, \cD \bfq)$, and 
corresponding actions
\be
A[\bfq; N] = \int_{t_1}^{t_2} dt\, N L(\bfq, \cD \bfq).
\label{1.4}
\ee
Under variations of the variables vanishing at the boundary, the variation of the action is
\be
\del A = \int_{t_1}^{t_2} dt \left[ N \del \bfq \cdot \lh \dd{L}{\bfq} - \cD \dd{L}{\cD\bfq} \rh +
 \del N \lh L - \cD\bfq \cdot \dd{L}{\cD\bfq} \rh \right].
\label{1.5}
\ee
Hence stationary points of the action must satisfy two conditions: first, 
the Euler-Lagrange equation must hold in the form
\be
\dd{L}{\bfq} - \cD \dd{L}{\cD \bfq} = 0, 
\label{1.6}
\ee
and second, there is a first-class constraint imposed by the variation with respect to the 
lapse function:
\be
L - \cD \bfq \cdot \dd{L}{\cD \bfq} = 0.
\label{1.7}
\ee
The left-hand side of this equation is seen to be the Legendre transform of the lagrangean
with respect to the physical degrees if freedom; therefore defining the covariant momentum 
and hamiltonian:
\be
\bfP = \dd{L}{\cD \bfq}, \hs{2} \cH = \bfP \cdot \cD \bfq - L,
\label{1.8}
\ee
the constraint (\ref{1.7}) is seen to amount to the vanishing of the hamiltonian: $\cH = 0$.
The implication of the reprametrization invariance is, that the phase space is reduced to the 
hypersurface where this condition holds. Note, that there is no canonical momentum associated
with the lapse function:
\be
p_N = \dd{L}{\dot{N}} = 0.
\label{1.8.1}
\ee 
Of course this just signifies that $N$ is a pure gauge degree of freedom, counting copies of 
the physical phase space related by different choices of the time parameter. The vanishing of 
the hamiltonian guarantees that the actual histories of the physical system do not depend on the 
specific choice. 

In the following we apply this formalism in particular to lagrangeans quadratic in the velocities:
\be
L = \frac{1}{2}\, G_{ij}(\bfq) \cD q^i \cD q^j - V(\bfq), 
\label{1.9}
\ee
where $G_{ij}(\bfq)$ is the metric on the configuration space; then 
\be
P_i = G_{ij}(\bfq)\, q^j, \hs{2} \cH = \frac{1}{2}\, G^{ij}(\bfq) P_i P_j + V(\bfq). 
\label{1.10}
\ee
Note that for positive definite metrics the hamiltonian constraint can be solved non-trivially only
in domains where $V < 0$; if $V = 0$ the only solutions allowed are constant ones: $ \cD \bfq = 0$, 
and for $V > 0$ no solutions can exist at all. Reparametrization invariance is more relevant for
theories (\ref{1.9}) with indefinite metrics. This is also the precondition for the existence
of internal clock variables. 

The hamiltonian constraint provides a convenient derivation of the Hamilton-Jacobi equation. 
Consider a solution of the constraint and Euler-Lagrange equation $\bfq_c(t)$ passing through 
the points $\bfq_1$ at time $t_1$ and $\bfq_2$ at time $t_2$. Then the action evaluated on this 
history becomes
\be
S(\bfq_2, \bfq_1) = \int_{t_1}^{t_2} dt N \ld L \right|_{\bfq_c} 
 = \int_{t_1}^{t_2} dt N \ld \cD \bfq \cdot \dd{L}{\cD \bfq} \right|_{\bfq_c} 
 = \int_{\bfq_1}^{\bfq_2} \ld d \bfq \cdot \bfP \right|_{\bfq_c}.
\label{1.11}
\ee
As a result 
\be
\dd{S}{\bfq_2} = \bfP_2, \hs{2} \dd{S}{\bfq_1} = - \bfP_1, \hs{2} \dd{S}{t_2} = \dd{S}{t_1} = 0. 
\label{1.12}
\ee
The absence of explicit dependence on the initial and final times $(t_1, t_2)$ reflects the 
hamiltonian constraint. Related to this is the gauge-independence of the Hamilton-Jacobi function:
\be
\ld \dd{S}{N} \right|_{\bfq_1, \bfq_2}= 0.
\label{1.13}
\ee
Eqs.\ (\ref{1.10}) and (\ref{1.12}) imply that $S$ satisfies the Hamilton-Jacobi equation 
\be
\frac{1}{2}\, G^{ij}(\bfq) \dd{S}{q^i} \dd{S}{q^j} + V(\bfq) = 0.
\label{1.14}
\ee
Observe, that this equation is manifestly gauge independent. 

\section{Pure FLRW cosmology}

An example of a reparametrization invariant theory is provided by the FLRW models of cosmology. 
In this section we consider the pure model without any additional degrees of freedom, such as 
cosmological scalar fields, but allowing for a cosmological constant. Taking natural units in which 
$8 \pi G = c = 1$, the FLRW metric reads
\be
ds^2 = - N^2 dt^2 + a^2 \lh \frac{dr^2}{1 - kr^2} + r^2 d\Og^2 \rh,
\label{2.1}
\ee
where $N(t)$ is the lapse function, $a(t)$ is the scale factor, and $k$ is the constant determining the 
spatial curvature: $k = 0$ for flat space, $k = +1$ for spherical space and $k = -1$ for
hyperbolic space. For this metric the Einstein action per unit co-ordinate volume for GR, including 
a cosmological constant, becomes
\be
A = \int_{t_1}^{t_2} dt N \lh - 3 a \lh \cD a \rh^2 + 3 k a- \ve a^3 \rh.
\label{2.2}
\ee
In these models $a(t)$ is the only physically relevant dynamical parameter. Its conjugate momentum is
\be
p_a = \dd{L}{\cD a} = - 6 a \cD a,
\label{2.3}
\ee
whilst the covariant hamiltonian defined according to (\ref{1.8}) reads
\be
\cH = \cD a \dd{L}{\cD a} - L = - \frac{p_a^2}{12a} - 3 ka + \ve a^3.
\label{2.4}
\ee 
Note, that with $a > 0$ the pure FLRW models actually have a negative-definite metric for 
the kinetic terms, indicating that $a$ is a potential internal clock variable. 

The expansion rate of the universe is expressed by the covariant Hubble parameter
\be
H = \frac{\cD a}{a} = - \frac{p_a}{6 a^2}.
\label{2.5}
\ee 
It can be used to simplify the cosmological evolution equations. 
In particular the hamiltonian becomes 
\be
\cH = a^3 \lh - 3 H^2 - \frac{3k}{a^2} + \ve \rh,
\label{2.6}
\ee
whilst the Euler-Lagrange equation takes the form
\be
2 \cD H + 3 H^2 + \frac{k}{a^2} - \ve = 0.
\label{2.7}
\ee
Using the hamiltonian constraint this can actually be written in a form which is valid for all $k$:
\be
\cD H + H^2 - \frac{\ve}{3} = 0.
\label{2.8}
\ee
Expression (\ref{2.4}) now provides the form of the Hamilton-Jacobi equation for pure FLRW
cosmology:
\be
-\frac{1}{12a} \lh \dd{S}{a} \rh^2 - 3 ka + \ve a^3 = 0.
\label{2.9}
\ee
Obviously, this equation can be solved only if $\ve a^2 - 3 k  \geq 0$. Upon this condition, 
we look for the solutions of this equation passing through the points $(a_1, a_2)$. These
solutions are given by 
\be
S(a_2, a_1) = - \frac{6}{\ve} \lh \frac{\ve a_2^2}{3} - k \rh^{3/2} + \frac{6}{\ve} \lh \frac{\ve a_1^2}{3} - k \rh^{3/2}.
\label{2.10}
\ee
It follows, that the momenta are
\be 
p_{a} =  - 6 a \sqrt{\frac{\ve a^2}{3} - k}, 
\label{2.11}
\ee
evaluated at either end point of the trajectory. The covariant Hubble parameter is: 
\be
H(a) = \sqrt{ \frac{\ve}{3} - \frac{k}{a^2}}. 
\label{2.12}
\ee
This is seen to reproduce the constraint for the hamiltonian (\ref{2.6}). At any time the rate of change of the 
scale factor, expressed by the Hubble parameter, depends only on the instantaneous value of $a$ itself, but 
not on the co-ordinate time at which this value is reached. This observation also explains why the 
Hamilton-Jacobi function (\ref{2.10}) is fully separable in initial and final terms. 

For any choice of time co-ordinate one can now relate the time to the value of the scale parameter, 
by integrating eq.\ (\ref{2.12}):
\be
\int_{t_1}^{t_2} dt N(t) = \sqrt{3}\, \int_{a_1}^{a_2} \frac{da}{\sqrt{ \ve a^2 - 3k }}.
\label{2.13}
\ee
The integral on the right-hand side is elementary and can be performed explicitly in all relevant cases.
We check that it reproduces the standard results. First, for $k = +1$ we have to require $\ve  a^2 > 3$; 
thus $\ve > 0$ and $a$ has a minimal value $\sqrt{3/\ve}$. If we identify $t = 0$ with the time at which 
the scale factor reaches the minimum, the functional form of $a(t)$ becomes
\be
a(t) = \sqrt{\frac{3}{\ve}}\, \cosh \lh \sqrt{\frac{\ve}{3}} \int_0^t N(t') dt' \rh.
\label{2.14}
\ee
Next in the spatially flat case $k = 0$ either $\ve = 0$, which yields Minkowski space with $H = 0$ 
and $a =$ constant; or $\ve > 0$, corresponding to de Sitter space and exponential expansion. Explicitly: 
\be
a(t) = a(0) e^{\sqrt{\ve/3}\, \int_0^t N(t') dt'}.
\label{2.15}
\ee
Finally, in hyperbolic space $k = -1$ all values of $\ve$ are allowed, with the restriction that for $\ve < 0$
the scale factor reaches a maximum value $\sqrt{3/|\ve|}$, beyond which the universe collapses again. 
In detail: 
\be
\ba{ll}
\mbox{for}\, \ve > 0: & \dsp{ a(t) = \sqrt{\frac{3}{\ve}}\, \sinh \lh \sqrt{\frac{\ve}{3}} \int_0^t N(t') dt' \rh;  }\\
 & \\
\mbox{for}\, \ve = 0: &\dsp{ a(t) = \int_0^t N(t') dt'; }\\
 & \\
\mbox{for}\, \ve < 0: &\dsp{ a(t) = \sqrt{ \frac{3}{|\ve|}}\, \sin \lh \sqrt{ \frac{|\ve|}{3}} \int_0^t N(t') dt' \rh. }
\ea
\label{2.16}
\ee
In all cases we have taken $t = 0$ to be the time of the beginning of expansion: $a(0) = 0$. 

\section{The cosmological $\sg$-model} 

In view of their invariance under reparametrizations, it is straightforward to include scalar fields in models 
for the evolution of homogeneous and isotropic universes. This is of considerable interest, not only 
because they can play a role in the development of the early universe, but also because of their role
in the microscopic physics of matter and interactions. As scalar fields play a part in determining the values 
of  fundamental parameters of physics, like the Fermi constant of weak interactions, these values can 
depend a priori on the history of the universe. 

The scalar fields most likely to be of interest in any scenario are those with long range and small masses, 
at least compared to the Planck scale which governs the very early universe. Certainly in the context of
particle physics this immediately raises the question as to what would guarantee the stability of these 
small masses. A particular scenario is the stabilization of light scalars by the Nambu-Goldstone mechanism 
which accompanies spontaneous symmetry breaking \ct{frieman-etal1995,cadamuro-redondo2011}. 
Another possibility for generating light scalars is the appearance of moduli after dimensional reduction 
from higher dimensions \ct{damour-henneaux2000,pioline-waldron2002,kleinschmidt-nicolai2010}. 
Light scalars of these types can be described effectively by $\sg$-models, and it is the cosmological 
evolution of such models that we consider in this section. 

In homogeneous and isotropic universes collective degrees of freedom can be carried by scalar fields 
$\vf^i(t)$ depending only on cosmic time $t$; this is any time co-ordinate for which the equal-time surface 
have constant curvature, describe by the FLRW space-time metric (\ref{2.1}). The action of such a set 
of fields coupled to the large-scale cosmological space-time reads
\be
A = \int dt N \left[ - 3 a (\cD a)^2 + \frac{1}{2}\, a^3\, G_{ij} \cD \vf^i \cD\vf^j + 3ka - \ve a^3 \right].
\label{3.1}
\ee 
Here $G_{ij}(\vf)$ is the metric on the scalar manifold, which admits Killing vectors for any non-linear 
(broken) and linear (non-broken) symmetries of the theory. In the limit of broken exact symmetries 
the fields are rigidly massless, and we omit any potential terms for the scalars. For light, but not 
massless, pseudo-Goldstone scalars our treatment needs to be slightly generalized, as discussed
in sect.\ \ref{s6} below. 

Starting from the action (\ref{3.1}) we derive the covariant momenta and hamiltonian: 
\be
p_a = \dd{L}{\cD a} = - 6 a^2 H, \hs{2} p_i = \dd{L}{\cD \vf^i} = a^3 G_{ij} \cD \vf^j,
\label{3.2}
\ee 
and
\be
\cH = - \frac{1}{12a}\, p_a^2 + \frac{1}{2a^3}\, G^{ij} p_i p_j - 3ka + \ve a^3.
\label{3.3}
\ee
In configuration space, the Euler-Lagrange equations and constraint take the form 
\be
\ba{l}
\dsp{ 2 \cD H + 3H^2 + \frac{1}{2}\, G_{ij} \cD \vf^i \cD \vf^j + \frac{k}{a^2} - \ve = 0, }\\
 \\
\dsp{ \cD^2 \vf^i + \Gam_{jk}^{\;\;\;i} \cD \vf^j \cD \vf^k + 3 H \cD \vf^i = 0, }\\
 \\
\dsp{ 3H^2 - \frac{1}{2}\, G_{ij} \cD \vf^i \cD \vf^j + \frac{3k}{a^2} - \ve = 0. }
\ea
\label{3.4}
\ee
The scalar field equation can be integrated to give 
\be
\cD \lh a^6 G_{ij} \cD \vf^i \cD \vf^j \rh = 0 \hs{1} \Rightarrow \hs{1} G_{ij} \cD \vf^i \cD \vf^j = \frac{\og^2}{a^6},
\label{3.5}
\ee
where we have assumed that the scalar metric $G_{ij}$ is positive definite and $\og^2$ is a constant 
of integration. Combining the remaining two equations can be transformed to 
\be
3H^2 = \frac{\og^2}{2a^6} - \frac{3k}{a^2} + \ve, \hs{2}
\cD H + 3H^2 + \frac{2k}{a^2} - \ve = 0.
\label{3.6}
\ee
It is straightforward to check that the first of these equations is an integral solution of the second one. 

In view of the hamiltonian (\ref{3.3}) the Hamilton-Jacobi equation for these models is:
\be
- \frac{1}{12a} \lh \dd{S}{a} \rh^2 + \frac{1}{2a^3}\, G^{ij} \dd{S}{\vf^i} \dd{S}{\vf^j} - 3ka + \ve a^3 = 0.
\label{3.7}
\ee
Now as the potential terms depend only on the scale factor, the equation can be separated:
\be
S(a, \vf^i) = S_{gr}(a) + S_{\sg}(\vf^i), 
\label{3.8}
\ee
such that 
\be
p_a = \dd{S_{gr}}{a}, \hs{2} p_i = \dd{S_{\sg}}{\vf^i}.
\label{3.9}
\ee
Then separation of variables is achieved by writing 
\be
G^{ij}\, \dd{S_{\sg}}{\vf^i} \dd{S_{\sg}}{\vf^j} = \og^2, \hs{2} 
\frac{a^2}{6} \lh \dd{S_{gr}}{a} \rh^2 + 6 ka^4 - 2\ve a^6 = \og^2,
\label{3.10}
\ee
where $\og^2$ is a constant of integration, representing the energy of the scalar fields. 
Observe, that the dynamics of  gravitational sector only depends on the $\sg$-model fields 
through this constant. The general solution for the gravitational terms in the Hamilton-Jacobi 
function is obtained by direct integration:
\be
S_{gr}(a_2, a_1) = - \sqrt{6}\, \og\, \int_1^2 \frac{da}{a} \sqrt{1 - \frac{6k}{\og^2} a^4 + \frac{2\ve}{\og^2} a^6}.
\label{3.11}
\ee
This integral can be performed explicitly in two special cases: \\
1.\ If $k = 0$ the solution reads
\be
S_{gr} = - \sqrt{\frac{2}{3}}\, \og \left[ \ln \lh \frac{a_2}{a_1} \rh^3 - 
 \ln \lh \frac{1 + \sqrt{1 + 2\ve a_2^6/\og^2}}{1 + \sqrt{1 + 2 \ve a_1^6/\og^2}} \rh 
 + \sqrt{1 + \frac{2\ve}{\og^2}\, a_2^6} - \sqrt{ 1 + \frac{2\ve}{\og^2}\, a_1^6} \right]; 
\label{3.12}
\ee
2.\ if $\ve = 0$ the solution for the various $k$-values is 
\be
S_{gr} = - \sqrt{\frac{3}{2}}\, \og \left[ \ln \lh \frac{a_2}{a_1} \rh^2 
 - \ln \lh \frac{1 + \sqrt{1 - 6ka_2^4/\og^2}}{1 + \sqrt{1 - 6ka_1^4/\og^2}} \rh  
 + \sqrt{1 - \frac{6k}{\og^2}\, a_2^4} - \sqrt{1 - \frac{6k}{\og^2}\, a_1^4} \right].
\label{3.13}
\ee
It follows, that the on-shell momenta are given by
\be
p_a = - 6 a^2 H, \hs{2} H = \frac{\og}{a^3 \sqrt{6}}\, \sqrt{1 - \frac{6k}{\og^2} a^4 + \frac{2\ve}{\og^2} a^6}. 
\label{3.14}
\ee
Following eq.\ (\ref{2.13}) the relation between the scale factor and the time co-ordinate can be determined
here as well: 
\be
\int_{t_1}^{t_2} N(t) dt = \sqrt{6}\, \int_{a_1}^{a_2} da\, \frac{a^2}{\sqrt{\og^2 - 6ka^4 + 2 \ve a^6}}.
\label{3.15}
\ee
For $k = 0$, and with the convention $a(0) = 0$, evaluation of the integral gives: 
\be
\ba{ll}
\mbox{for}\, \ve > 0: & \dsp{ a^3(t) = \frac{\og}{\sqrt{2\ve}}\, \sinh  \lh \sqrt{3\ve} \int_0^t N(t') dt' \rh; }\\
 & \\
\mbox{for}\, \ve = 0: & \dsp{ a^3(t) = \sqrt{\frac{3}{2}}\, \og \int_0^t N(t') dt'; }\\
 & \\
\mbox{for}\, \ve < 0: & \dsp{ a^3(t) = \frac{\og}{\sqrt{2|\ve|}}\, \sin \lh \sqrt{3|\ve|} \int_0^t N(t') dt' \rh, }
\ea
\label{3.16}
\ee
where in the last case for single-valuedness we have to impose the restriction 
\be
0 < \sqrt{3|\ve|} \int_0^t N(t') dt' < \pi. 
\label{3.17}
\ee

\section{Solving for the scalar fields: the $O(N)$ $\sg$-model} 

The first equation (\ref{3.10}) is the Hamilton-Jacobi equation for the scalar fields of the $\sg$-model:
\[
G^{ij}\, \dd{S_{\sg}}{\vf^i} \dd{S_{\sg}}{\vf^j} = \og^2. 
\]
In order to make progress we have to specify the nature of the scalar field metric. An explicit example
is provided by the $O(N)$ $\sg$-model \ct{davis-macfarlane-vholten1983}, in which a scalar field $\bFg$ 
with the free action 
\be
A_{\sg} = \frac{1}{2}\, \int_1^2 dt N a^3 \lh \cD \bFg \rh^2.
\label{4.1}
\ee
takes values on the sphere $S^{N-1}$: 
\be
\bFg^2 = 1, \hs{2} \bFg = (\Fg^1, ..., \Fg^N).
\label{4.2}
\ee
This constraint can be solved by introducing projective co-ordinates
\be
\ba{l}
\dsp{ \Fg^i = \frac{\vf^i}{\sqrt{ 1 + \uvf^2 }}, \hs{2} i = (1, ..., N-1), }\\
 \\
\dsp{ \Fg^N = \frac{1}{\sqrt{ 1 + \uvf^2 }}. }
\ea
\label{4.3}
\ee
In this parametrisation the action takes the form (\ref{3.1}) with
\be
G_{ij} = \frac{\del_{ij}}{1 + \uvf^2} - \frac{\vf_i \vf_j}{(1 + \uvf^2)^2}, \hs{2}
G^{ij} = \lh 1 + \uvf^2 \rh \lh \del^{ij} + \vf^i \vf^j \rh.
\label{4.4}
\ee
The components of the connection are
\be
\Gam_{ij}^{\;\;\;k} = - \frac{\del_i^k\, \vf_j + \del_j^k\, \vf_i}{1 + \uvf^2}.
\label{4.4.1}
\ee
Then the Hamilton-Jacobi equation for the scalar fields takes the explicit form 
\be
\lh \dd{S_{\sg}}{\uvf} \rh^2 + \lh \uvf \cdot \dd{S_{\sg}}{\uvf} \rh^2 = \frac{\og^2}{1 + \uvf^2}.
\label{4.5}
\ee
As shown below, an explicit solution of this equation is
\be
S_{\sg} = \og \arctan \sqrt{\uvf^2},
\label{4.6}
\ee
with the momentum expressed by
\be
\up = \dd{S_{\sg}}{\uvf} = \frac{\og \un}{1 + \uvf^2}, \hs{2} \un = \frac{\uvf}{\sqrt{\uvf^2}}.
\label{4.7}
\ee
Now by inverting the second equation (\ref{3.2}) one gets
\be
\cD \vf^i = \frac{1}{a^3} G^{ij} p_j = \frac{\og}{a^3} \lh 1 + \uvf^2 \rh \frac{\vf^i}{\sqrt{\uvf^2}} \hs{1} 
 \Rightarrow \hs{1} \frac{\uvf \cdot \cD\uvf}{\uvf^2} = \frac{\og}{a^3} \lh 1 + \uvf^2 \rh \frac{1}{\sqrt{\uvf^2}},
\label{4.8}
\ee
which implies 
\be
\cD \un = \frac{1}{\sqrt{ \uvf^2}} \lh \cD \uvf - \frac{\uvf \cdot \cD \uvf }{\uvf^2}\, \uvf \rh 
  = 0.
\label{4.9}
\ee
Hence the unit vector $\un$ is constant, and $\uvf$ can change its length, but not its direction:
\be
\uvf(t) = f(t) \un \hs{1} \Rightarrow \hs{1} f(t) = \un \cdot \vf(t).
\label{4.10}
\ee
It follows, that 
\be
\cD f = \frac{\og}{a^3} \lh 1 + f^2 \rh \hs{1} \Rightarrow \hs{1} \cD \arctan f = \frac{\og}{a^3}.
\label{4.11}
\ee
However, in order to write the solution for $\vf$ in a form which is independent of the choice of
time parameter, one can replace the dependence on $t$ by the dependence on $a$:
\be
\frac{d}{da} \arctan f = \frac{\og}{a^4 H} = \frac{\og \sqrt{6}}{a \sqrt{\og^2 - 6ka^4 + 2 \ve a^6 }}. 
\label{4.12}
\ee
This can be integrated for the known cosmological solutions for the scale factor. 
For example, in flat FLRW space-time ($k = 0$) one gets \\
\be
f(a_2) = \tan \left\{ \sqrt{\frac{3}{2}} \left[ \ln \lh \frac{a_2}{a_1} \rh^3 - \ln \lh \frac{1 + \sqrt{1 + 2\ve a_2^6/\og^2}}{
 1+ \sqrt{1 + 2\ve a_1^6/\og^2}} \rh \right] + \arctan f(a_1) \right\},
\label{4.13}
\ee
which is equivalent to 
\be
\frac{f(a_2) - f(a_1)}{1 + f(a_2) f(a_1)} = \tan \sqrt{\frac{2}{3}} \left[ \ln \lh \frac{a_2}{a_1} \rh^3 
 - \ln \lh \frac{1 + \sqrt{1 + 2\ve a_2^6/\og^2}}{1 + \sqrt{1 + 2\ve a_1^6/\og^2}} \rh \right].
\label{4.14}
\ee
Clearly, a convenient initial condition is to take $a_1$ the scale at which $f(a_1) = 0$. Note that to 
avoid singularities, the solution requires the domain of the scale parameter to be restricted to
\be
- \frac{\pi }{2} \sqrt{\frac{3}{2}} < \ln \lh \frac{a_2}{a_1} \rh^3 
 - \ln \lh \frac{1 + \sqrt{1 + 2\ve a_2^6/\og^2}}{1 + \sqrt{1 + 2\ve a_1^6/\og^2}} \rh < \frac{\pi}{2} \sqrt{\frac{3}{2}}.
\label{4.15}
\ee

\section{Massive scalars \label{s6}}

The cosmological models with massless scalars can be extended with mass terms, as happens for example
due to soft breaking of the symmetries that generate Goldstone bosons. Assuming the masses to be degenerate, 
the action (\ref{3.1}) is generalized to
\be
A = \int dt N a^3 \left[ - 3 \lh \frac{\cD a}{a} \rh^2 + \frac{3 k}{a^2} - \ve + \frac{1}{2} G_{ij} \cD \vf^i \cD \vf^j
 - \frac{1}{2} m^2 \vf_i^2  \right].
\label{m1.1}
\ee
The corresponding Hamilton-Jacobi equation reads
\be
 \frac{1}{12 a} \lh \dd{S}{a} \rh^2 - 3k a + \ve a^3 + \frac{1}{2a^3} G^{ij} \dd{S}{\vf^i} \dd{S}{\vf^j} 
  + \frac{1}{2}\, m^2 a^3 \vf_i^2 = 0.
\label{m1.2}
\ee
This equation is no longer separable in terms of scale factor and scalar fields. However, assuming the 
mass $m$ to be small on the Planck scale, we can construct perturbative solutions by expanding $S$ 
in the form
\be
S = S_{gr} + S_{\sg} + m^2 S_1,
\label{m1.3}
\ee
To first order in $m^2$ the perturbation $S_1$ must satisfy
\be
a H \dd{S_1}{a} + \cD \vf^i \dd{S_1}{\vf^i} + \frac{1}{2}\, a^3 \vf_i^2 = 0,
\label{m1.4}
\ee
Here $(a, H, \vf^i)$ are taken to represent the unperturbed solutions of the massless theory constructed 
above. In the context of the $O(N)$-model we have been able to write these solutions in time-independent 
form (\ref{4.12}), (\ref{4.13}). We can use the same procedure to write eq.\ (\ref{m1.4}) as
\be
H \lh \dd{S_1}{a} + \frac{d\vf^i}{da} \dd{S_1}{\vf^i} \rh + \frac{1}{2}\, a^2 \vf_i^2  = 0,
\label{m1.6}
\ee
For the $O(N)$-model an explicit solution can be constructed by parametrizing $S_1$ in the form
\be
S_1 = \sum_{k \geq 0} \sg_k(a) f^k, \hs{2} f^2 = \uvf^2.
\label{m2.1}
\ee
Using the result (\ref{4.12}) and introducing the comoving volume $\nu = a^3$, this equation takes the form
\be
\sum_{k \geq 0} \left[ 3 H \frac{d\sg_k}{d\nu} + \frac{\og (k+1) \sg_{k+1}}{\nu^2} \lh 1 + f^2 \rh \right] f^k = - \frac{f^2}{2}, 
\label{m2.4}
\ee
and by comparing powers of $f$ on both sides of the equation one gets 
\be
\ba{ll}
\dsp{ 3 H \frac{d\sg_0}{d \nu} + \frac{\og \sg_1}{\nu^2} = 0, }& 
\dsp{ 3 H \frac{d\sg_1}{d \nu} + \frac{2 \og \sg_2}{\nu^2} = 0, }\\
 & \\
\dsp{ 3 H \frac{d\sg_2}{d \nu} + \frac{\og (3 \sg_3 + \sg_1)}{\nu^2} = - \frac{1}{2}, }& 
\dsp{ 3 H \frac{d\sg_3}{d \nu} + \frac{\og (4 \sg_4 + 2 \sg_2)}{ \nu^2} = 0, }\\
\ea
\label{m2.5B}
\ee 
Here we only consider the case of flat background space-time $k = \ve = 0$, with
\[
H = \frac{\og}{\nu \sqrt{6}}.
\]
The solution of (\ref{m2.4}), (\ref{m2.5B}) then is
\be
\sg_k = \frac{1}{2\og}\, \ag_k\, \nu^2,
\label{m2.9}
\ee
where the coefficients $\ag_k$ satisfy the relations
\be
\ba{ll}
\sqrt{6} \ag_0 + \ag_1 = 0, & \sqrt{6} \ag_1 + 2 \ag_2 = 0, \\
 & \\ 
\sqrt{6} \ag_2 + 3 \ag_3 + \ag_1 = - 1, & \sqrt{6} \ag_3 + 4 \ag_4 + 2 \ag_2 = 0, \\
\ea
\label{m2.10}
\ee 
These equations are solved by
\begin{equation}
\alpha_1 = - \sqrt{6} \alpha_0, \; \; \; \alpha_2 = 3 \alpha_0, \; \; \; \alpha_3 = - \frac{1}{3} - \frac{2}{3}\, \sqrt{6}\, \alpha_0,
\label{m2.12B}
\end{equation}
for the first four coefficients, and a recurrence relation
\be
\ag_{k+1} = - \frac{\sqrt{6}}{k+1}\, \ag_k - \lh \frac{k-1}{k+1} \rh \ag_{k-1},
\label{m2.12}
\ee
for the remaining coefficients. The final expression for $S_1$ now takes the simple form 
\be
m^2 S_1 = \frac{m^2 a^6}{2\og} \sum_{k \geq 0} \ag_k f^k + {\cal O}(m^4)
\label{m2.11}
\ee
The series defined by ({\ref{m2.11}) is absolutely convergent for $\mid f \mid <1$,  because the second
term of the recurrence relation (\ref{m2.12}) tends to $1$ when $k \rightarrow \infty$.  This of course
restricts our solutions to small perturbations around the equilibrium solution.  

In particular, for the case $\ag_0 = 0$ we get 
\be
\ag_1 = \ag_2 = 0, \hs{1} \ag_3 = - \frac{1}{3}, \hs{1} \ag_4 = \frac{\sqrt{6}}{12}, \hs{1} .... 
\label{m3.1}
\ee 
The equation for the second-order contribution $S_2 (a, \varphi)$ is also easily obtained:
\be
- \frac{1}{6} \, \left( \frac{\partial S_0}{\partial a} \right) \left( \frac{\partial S_2}{\partial a} \right)
+ \frac{1}{a^3} \,  G_{ij} \, \left( \frac{\partial S_0}{\partial \varphi^i} \right) \left( \frac{\partial S_2}{\partial \varphi^j} \right)
= \frac{1}{12} \left( \frac{\partial S_1}{\partial a} \right)^2 + 
\frac{1}{a^3} G_{ij} \left( \frac{\partial S_1}{\partial \varphi^i} \right) \left( \frac{\partial S_1}{\partial \varphi^j} \right).
\label{m2.14B}
\ee
We see that as usual, the left-hand side containing the unknown function $S_2$ is of the same form as the linear 
approximation for $S_1$, whereas the right-hand side contains expressions quadratic in $S_1$, which is a known 
function due to the previously found solution. The initial condition has been already set, so here the intial condition 
should be $S_2 = 0$ at $a=0$.

\section{Discussion} 

In this paper we have developed the Hamilton-Jacobi formalism for theories with time-reparametrization 
invariance. These theories have non-trivial solutions only in the presence of negative-energy contributions
to the covariant hamiltonian. Using covariant definitions of the generalized momenta, solutions can be 
obtained at once for all choices of gauge. The formalism was applied to models of the cosmological 
evolution of anisotropic and homogeneous universes of the FLRW type, allowing 
for extensions with spatiallly homogeneous scalar fields.  In particular we studied solutions for theories 
with Goldstone fields, as parametrized by non-linear $\sg$-models. In these theories the Hamilton-Jacobi 
function is separable, allowing for explicit solutions for the gravitational sector of the models. The case 
of the non-linear $O(N)$-models was considered in more detail and an explicit solution was obtained. 
This solution actually represents motion with constant {\it internal} angular momentum on a great circle 
on $S^{N-1}$ passing through the poles. In addition, we showed how to construct solutions for scalar 
fields with non-vanishing mass, by  treating the mass terms as a perturbation of the massless theory.  

Our solution of the massless $O(N)$-model, although representative of the general solution, is not unique, 
as the symmetries of the models allow one to rotate the solution to any great circle obtained by cutting the 
sphere with a plane passing through the center. In fact, as the choice of orientation of $\un$ in (\ref{4.10}) 
is free, all great circles passing through the poles are obtained by rotations over a fixed angle of $\uvf$, or 
equivalently $\un$, in the $(N-1)$-dimensional plane. More solutions, corresponding to great circles in 
different planes rotated w.r.t.\ the polar axis, are generated by the modified Hamilton-Jacobi function
\be 
S'_{\sg} = S_{\sg} + \del S_{\sg}, \hs{2} \del S_{\sg} = \og\, \frac{\ueps \cdot \uvf}{\sqrt{\uvf^2}}
 = \og \ueps \cdot \un, 
\label{5.1}
\ee
to first order in the $N-1$ parameters $\ueps$. Indeed, this modification transforms the scalar field 
conjugate momentum to 
\be
\up' = \up + \del \up, \hs{2} \del \up = \frac{\og}{\sqrt{\uvf^2}} \lh \ueps - \un \un \cdot \ueps \rh,
\label{5.2}
\ee
while leaving the value of $\og^2$ unchanged:
\be
\up \cdot \del \up = \uvf \cdot \del \up = 0.
\label{5.3}
\ee
Of course, our conclusions are modified in the quantum regime, where the values of $\og^2$ are 
limited to 
\[
\og^2 \hs{1} \rightarrow \hs{1} l(l+N-2), \hs{1} l = 0,1,2,...
\]
and only a finite number of orientations of the internal angular momentum are allowed.
It is to be noticed that the models defined here possess a dynamical $SO(2,1)$ invariance 
\ct{pioline-waldron2002}, which can be used to construct the spectrum. 
\vs{2}

\nit
{\bf Acknowledgement} \\
The work reported here is part of the research programme of the Foundation for Research of Matter (FOM). 
The authors are indebted to Giuseppe D'Ambrosi for useful discussions.

\end{document}